\documentclass[showpacs, twocolumn, aps, prb]{revtex4-1} 

        \expandafter\let\csname equation*\endcsname\relax
        \expandafter\let\csname endequation*\endcsname\relax
        \usepackage{amsmath,bm}
        \usepackage{amsfonts} 
        \usepackage{graphicx}
        \usepackage{url}
        \usepackage{hyperref}
        \usepackage{breakurl}   
        \usepackage[switch,columnwise]{lineno}  
        \usepackage{color}
        \usepackage{epstopdf}  
        
        \newcommand{\comment}[1]{}
    
            \makeatletter
            \newsavebox{\@brx}
            \newcommand{\llangle}[1][]{\savebox{\@brx}{\(\m@th{#1\langle}\)}%
              \mathopen{\copy\@brx\kern-0.5\wd\@brx\usebox{\@brx}}}
            \newcommand{\rrangle}[1][]{\savebox{\@brx}{\(\m@th{#1\rangle}\)}%
              \mathclose{\copy\@brx\kern-0.5\wd\@brx\usebox{\@brx}}}
            \makeatother        

\newcommand{\JL}{${\mathcal{J}}$ } 
\newcommand{\vbspi}{VBS(comm.) }

\newcommand{\qlropi}{QLRO }
\newcommand{\qlropihalf}{decoupled }
\newcommand{\MG}{Majudar-Ghosh }

\usepackage{bibentry}

\newcommand{\ignore}[1]{}
\newcommand{\bibfnamefont}[1]{#1}
\newcommand{\bibnamefont}[1]{#1}

\mathchardef\mhyphen="2D    

\newcommand{\jj}{$J_1\textrm{-}J_2$ }

\begin{document}


        \title{Matrix product state approach to a frustrated spin chain with long-range interactions}
        
    \author{Zhi-Hua Li and An-Min Wang}
    \address{Department of Modern Physics, University of Science and Technology of China, Hefei 230026, China}

    \begin{abstract}
        We make extensive simulations over a spin chain model that combines the
        frustrated $J_1\textrm{-}J_2$ spin chain and the long-range nonfrustrated
        $(-1)^{(r-1)}r^{-\alpha}$ decay interactions through the variational matrix product
        state method for both finite and infinite lengths. We study both the
        ground state entanglement and phase diagram. 
        We find that it is most entangled in the rotation invariant long-range
        ordered antiferromagnetic phase, where the entanglement scales
        approximately logarithmically.   
        We determine the development of the Majudar-Ghosh point to a disorder
        line from entanglement. And we determine approximately the transition from
        the dimerized and  incommensurate phase of the $J_1\textrm{-}J_2$ model to a
        decoupled phase by studying spin correlation and the dimerization order
        parameter.  
        Some implications for entanglement in systems with long-range interactions are stated. 

        \end{abstract} 


    \maketitle
    \modulolinenumbers[5]

\section{introduction}
    Quantum spin chains are the fertile ground to study strongly correlated
    quantum many body effects.  
    \comment{Besides, the have various practical realizations.}
    One of the most studied models is the quantum Heisenberg model. 
    \comment{
        It is exactly solvable by the Bethe Ansatz and well described by field
        theory. }
    For antiferromagnetic coupling ($J_1>0$) the spin correlation  of its
    ground state  decays as $~1/r$ up to a logarithm correction
    \cite{affleck1989critical,*giamarchi1989correlation} displaying quasi-long range order (QLRO). 
    The model with a next-nearest neighbour added is known as the \jj or
    zigzag spin chain. It is frustrated  when $J_2>0$. 
    A dimerization transition occurs at $J_2/J_1\simeq 0.2411$ after which it
    has  a valence-bond-solid (VBS) order
    \cite{okamoto1992fluid,eggert1996numerical} and incommensurate spiral spin
    correlation emerges after the \MG (MG) point at  $J_2/J_1=0.5$
    \cite{majumdar1969next,bursill1995numerical,white1996dimerization}. 
    \comment{
        As $J_2$ increases from 0, a spin gap opens at $J2_c=0.2411$
        \cite{okamoto1992fluid,eggert1996numerical} and a fluid-dimer transition occurs.
        }
    
    Beyond the next-nearest terms, the system can be built up with even
    long-range interactions (LRI). 
    The models with power law
    decay of LRI coupling $J_r\sim (-1)^{r-1}r^{-\alpha}$ have
    attracted much attention \cite{yusuf2004spin}, from which, intriguingly 
    true long-range ordered antiferromagnet (AFM) can be formed for small 
    enough $\alpha$ \cite{laflorencie2005critical}, even though it have been strictly
    ruled out from 1D short-range rotation invariant models at even zero temperature.  
    \comment{
        It is frustration free and its ground state
        have been calculated accurately by the quantum Monte Carlo algorithm
        \cite{laflorencie2005critical}. }
    
    Recently, Sandvik proposed the combination of the \jj model and the
    long-range nonfrustrated terms for studying the interplay between them
    \cite{sandvik2010ground}. The Hamiltonian is 
    \begin{equation}
        \begin{split}
        \label{eq:ham}
        &\qquad H = \sum\limits_{i = 1}^N  {\sum\limits_{j = i+1}^{N} {J_{|j-i|}} {{{\vec
        S}_i}\cdot{{\vec S}_j}} }  \\
        &J_2=g, \quad\quad {J_{r \ne 2}} = {{{{( - 1)}^{r - 1}}}}{{r^{-\alpha
        }}}/n(N,\alpha),
        \end{split}
        \end{equation}
    where the normalization factor $n(N, \alpha)\equiv{ {1 + \sum_{r = 3}^{N/2}
    {{{r^{-\alpha }}}} } }$  signifies that the sum of all the interactions
    (excluding $r=2$) on one site add to unity, and it also ensures finite
    energy per site for infinite $N$
    when $\alpha\le1$.  
    The ground state phase diagram has been investigated by Sandvik
    \cite{sandvik2010ground} using the exact diagonalization (ED) method, and
    later by Kumar and Soos \cite{kumar2013decoupled} using ED and other
    auxiliary methods. 
    \comment{
        Sandvik studied the ground state phase diagram through ground state or
        excitation state level crossings using the exact diagonalization (ED) method,
        and successfully determined several phase boundaries, especially the
        development of the dimerization transition point and then to a first order
        phase transition. But it is insufficient to give a reliable phase boundary
        at moderate frustration, due to restriction on the size. Kumar and Soos
        later studied the model also use ED and several auxiliary methods in
        particular the case of large $g$ regime. They give a phase boundary for
        moderate $g$ qualitatively different with Sandvik.  Note this regime should
        be should be related to the spiral state in the ab initio study of realistic metallic chains
        \cite{tung2011ab}.
        }
    Several phase boundaries have been accurately determined
    \cite{sandvik2010ground}, but it is still controversial in the regime with
    moderate frustration \cite{sandvik2010ground,kumar2013decoupled}, where it
     should be related to the spiral state in the ab initio study of realistic
    metallic chains \cite{tung2011ab}. Besides, the ground state entanglement
    has not been considered in both works. 
    In this paper, we restudy the model \ref{eq:ham} using the matrix product
    state (MPS) approach for the ground state entanglement and phase diagram.  
    As is established in recent years, entanglement enriches the
    characterization of quantum phases and phase transitions
    \cite{amico2008entanglement}. 
    Particularly in this model, it
    displays sudden drop along a first order phase transition.  And a line
    segment of minimum entanglement marks the development of the \MG point. 
    This method also entails evaluation of spin
    correlation accurately for several hundred sites, so that we can directly
    demonstrate the incommensurate behaviour in the regime of moderate
    frustration. And we improved the phase boundary in this
    regime, which should resolve the controversy among the previous works. 
     
    \comment{
        Another motivation of our work concerns the scaling of the entanglement
        entropy with subsystem size $L$ in the presence of LRI. In the last decade,
        many efforts have been put into the study of scaling of EE with subsystem
        sizes \cite{amico2008entanglement}. Especially an area law
        \cite{srednicki1993entropy,eisert2010colloquium}  of entanglement is
        conjectured, but not generally proved (except at 1D
        \cite{hastings2007area}), which states that EE scales as $L^{d-1}$, with $d$
        the dimension of the system.  This is remarkable, as it means that physical
        ground states are ``slightly'' entangled.  In particular for 1D systems it
        have been proved rigorously for gapped Hamiltonian \cite{hastings2007area},
        and is shown to be violated mildly by logarithmic divergence for gapless
        systems \cite{vidal2003entanglement, * calabrese2004entanglement, *
        holzhey1994geometric}. Whereas it is still not very clear  up to now the
        behaviour of entanglement in systems with LRI
        \cite{cirac2009renormalization}, especially that should then the area law
        be severely violated?  Recently Koffel et. al. \cite{koffel2012entanglement}
        made an achievement in this direction, showing interestingly that in the LRI
        transverse Ising model a gapped phase can even have logarithmic scaling of
        entanglement.
        }
        
    Another motivation of our work concerns the scaling of the entanglement
    entropy  with subsystem size in quantum many body
    systems. In the last decade, 
    much effort has been devoted to this problem 
    \cite{amico2008entanglement}. Especially an area law
    \cite{srednicki1993entropy, *eisert2010colloquium}  of entanglement is
    conjectured, which states that for a local and gapped
    Hamiltonian the ground state entanglement entropy of a subsystem scales
    as the boundary area.  This is remarkable, as it
    means that physical ground states are ``slightly'' entangled.  
    In particular for 1D systems it
    have been proved rigorously for gapped Hamiltonian \cite{hastings2007area},
    and is shown to be violated mildly by logarithmic divergence for gapless and
    conformal invariant systems \cite{vidal2003entanglement, * calabrese2004entanglement, *
    holzhey1994geometric}. 
    
    The area law is most generally attributed to the interactions being local. 
    The entanglement structure in LRI systems is less known
    \cite{cirac2009renormalization}.  Especially, one wonders that should then
    the area law be severely violated?  Some progresses have been made  
    in spin chains with LRI. In \onlinecite{latorre2005entanglement}, it was
    shown that  for the Lipkin-Meshkov-Glick model, which resembles
    the XY model but with infinite range interactions, the entanglement entropy
    scales at most logarithmically.  Lately,
    Koffel et. al.  \cite{koffel2012entanglement} studied  a transverse Ising
    model with power law decay LRI, and showed interestingly that a gapped phase can
    even have logarithmic scaling of entanglement.  That model is polarized, if
    the system is rotation invariant, stronger quantum fluctuation and
    entanglement will present, which to our knowledge has not
    been considered so far.  The model \ref{eq:ham} being rotation invariant is
    ideal for addressing this question. Besides, with a tunable frustrating
    term, it facilitates to examine when LRI takes effect in increasing of
    entanglement. We find that it is most entangled in the AFM long-range
    ordered phase. But, remarkably, the scaling of entanglement can be still
    fitted approximately with logarithm functions. This indicates that, in
    contrary to one might expected, the area law should be not severely violated in
    this system.    

    This work is organized as follows. In Sec \ref{sec:method} we briefly
    introduce the numerical methods. Sec.\ref{sec:results} presents the simulation
    results. In Sec. \ref{sec:results:phase:diagram} we show
    the ground state phase diagram, half chain entanglement
    entropy, and the development of the \MG point. 
    Some numerical pitfalls and partial solution to them are also manifested. 
    In Sec.\ref{sec:results:EE_scaling} we discuss the scaling of entanglement
    with subsystem sizes. In Sec.\ref{sec:resuts:moderate_g} we study the phase
    boundary at moderate frustration. Finally,  we conclude in
    Sec.\ref{sec:conclusion} with some implications stated.

\section{method} \label{sec:method}
    
    \comment{
        It is conventionally difficult to simulate systems with LRI using density
        matrix renormalization group (DMRG) \cite{white1992density, *white1993density} or MPS
        alike methods.  Crosswhite et. al. \cite{crosswhite2008applying} have made a
        major progress for LRI systems with power law type decay by approximately encoding 
        the Hamiltonians with matrix product operators (MPO), so that the computation cost is reduced
        considerably.
        We use this long range MPO method to represent  
        Eq.\ref{eq:ham} for finite and infinite systems 
        \footnote {
            In Eq.\ref{eq:ham}, $n(\alpha,N)$ diverges for infinite $N$ when
            $\alpha\leq 1$, which seems impossible to represent numerically. But
            in the long range MPO method, the power
            function coupling $P(r)=r^{-\alpha}$ is actually approximated with a $P'(r)$ by
            the sum of several exponential functions and  $P'(r)$ decays exponentially
            in large distance. So $\sum_{r } {{{P'(r)}}}$ [and also $n(\alpha,N)$]
            do converge in practice. } 
        in the parameter region of $(\alpha, g) \in [0.7, \infty)\times[0, 1.0]$ (we
        refer to \onlinecite{crosswhite2008applying} and
        \onlinecite{frowis2010tensor}
        for details), while with minor adaption for the $J_2$ term. 
        }
    
    It is conventionally difficult to simulate systems with LRI using density
    matrix renormalization group (DMRG) \cite{white1992density, *white1993density} or MPS
    alike methods.  Crosswhite et. al. \cite{crosswhite2008applying} have made a
    major progress for LRI systems with power law type decay by approximately encoding 
    the Hamiltonians with matrix product operators (MPO), so that the computation cost is reduced
    considerably.
    We use this long range MPO method to represent  
    Eq.\ref{eq:ham} for finite and infinite systems 
    \footnote {
        In Eq.\ref{eq:ham}, $n(\alpha,N)$ diverges for infinite $N$ when
        $\alpha\leq 1$, which seems impossible to represent numerically. But
        in the long range MPO method, the power
        function coupling $P(r)=r^{-\alpha}$ is actually approximated with a $P'(r)$ by
        the sum of several exponential functions and  $P'(r)$ decays exponentially
        in large distance. So $\sum_{r } {{{P'(r)}}}$ [and also $n(\alpha,N)$]
        do converge in practice. } 
    in the parameter region of $(\alpha, g) \in [0.7, \infty)\times[0, 1.0]$ (we
    refer to \onlinecite{crosswhite2008applying} and
    \onlinecite{frowis2010tensor}
    for details).  A minor adaption for the $J_2$ term is needed. We write Eq.\ref{eq:ham} in
    an equivalent form, 
    \begin{eqnarray}
        \label{eq:equiv}
     \begin{array}{l}
    H = {H^a} + {H^b}\\
    {H^a} = \frac{1}{{n(N,\alpha )}}\sum\limits_{i = 1}^N {\sum\limits_{j = i +
    1}^N {{{( - 1)}^{|j- i| - 1}}|j - i{|^{ - \alpha }}{S_i}{S_{j}}} } \\
    {H^b} = (g + \frac{1}{{{2^\alpha }n(N,\alpha )}})\sum\limits_{i = 1}^{N - 2}
    {{S_i}{S_{i + 2}}}. 
    \end{array}
    \end{eqnarray}
    Now $H^a$ has uniformly power law decay interaction (with alternation signs),
    so that the long range MPO method is applicable. The short-ranged $H^b$ can
     also be encoded in a MPO easily. Then the entire Hamiltonian is the sum of the
    two MPOs\cite{schollwock2011density}.

    We use the variational MPS (VMPS) algorithm \cite{dukelsky1998equivalence,*verstraete2004density,
    *mcculloch2007from, *schollwock2011density}  to simulate the ground states for
    finite open chains  with $N$ ranging from 16 to 100 and  truncation
    dimension $D$ up to 520.  It is
    implemented with the 1-site algorithm and density matrix correction
    \cite{white2005density} that reduces the chance being stuck.  The quality of
    the variational ground state is gauged by the average variance $v=(\langle
    H^2\rangle-\langle H \rangle^2)/N$, kept smaller than 1e-4 for the hardest
    case.  The iDMRG  algorithm  
    \cite{mcculloch2008infinite} (not exactly the conventional infinite size
    DMRG by White)
    is used to study infinite systems. This method exploits the
    translation invariance, such that  the computation effort is reduced and
    boundary effect is avoided. It is implemented with a 4-site unit cell, from
    which an infinite MPS (IMPS) representation can be reconstructed after convergence
    \cite{rommer1997class} for measuring physical observables.  The maximal $D$
    used
    is 1000 for generating a well
    converged fixed point  with truncation error restricted to smaller
    than 1e-6, while at some point we also use iDMRG to  generate a finite
    open chain with  even larger $D$.

    For a ground state wave function $|\psi\rangle$ on a finite open chain $[1,N]$  of
    length $N$, divided into a subsystem  $[1,L]$ and its
    complement $[L+1,N]$, we measure the entanglement entropy 
    \begin{equation}
        \label{eq:ee} 
        {S}=- {\rm{Tr(}}{\rho _L}\ln ({\rho _L})), 
        \end{equation}
    with $\rho_L=\text{Tr}_{ [L+1,N]}(|\psi\rangle\langle\psi|)$ the reduced
    density matrix for the subsystem. We also measure other quantities that will
    be defined later.

\section{simulation results}  \label{sec:results}
\subsection{ground state entanglement and phase diagram} \label{sec:results:phase:diagram}

        \begin{figure}   
              \centering
              \scalebox{0.55}[0.55]{\includegraphics{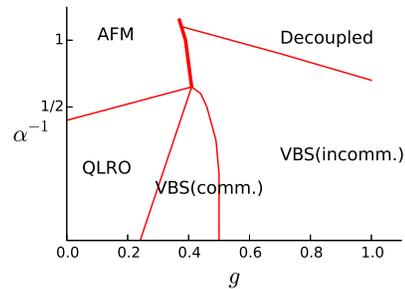}}
              \caption{\label{fig:phase:diagram} Ground state phase diagram for
              Eq.\ref{eq:ham}.  At $\alpha=\infty$, the model degenerates
              to the \jj chain, whose phases undergoes phase transitions under
              LRI with decreasing $\alpha$: the QLRO phase to a
              long range order AFM phase
              \cite{sandvik2010ground,laflorencie2005critical}; the commensurate
              VBS phase terminating at a
              multi-critical point at around $\alpha=1.7$ and $g=0.41$ and
              continued by a line of first order transition (thick solid line)
              \cite{sandvik2010ground}; the incommensurate VBS phase to a phase 
              decoupled for odd and even sublattice. 
              }
              \end{figure}
        \begin{figure}   
              \centering
              \scalebox{0.55}[0.55]{\includegraphics{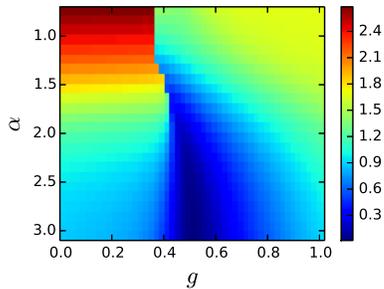}}
              \caption{\label{fig:EE} Distribution of the half chain entanglement
              entropy $S_{N/2}$ on the parameter domain $(\alpha,g) \in [0.7, 3.0]\times[0,1.0]$
              divided into $24\times50$ points for system size $N=60$. }
              \end{figure}
        We plot a schematic ground state phase diagram in Fig. \ref{fig:phase:diagram}   
        and show the distribution of entanglement on the parameter plane in
        Fig.\ref{fig:EE}. 
        The entanglement is generally higher as $\alpha$ reduces; 
        rather high in the top left corner while low in the bottom center. 
        These already give a rough profile of several of the phase boundaries.  
        \comment{
            The name of decoupled phase is the same as Kumar and Soos
            \cite{kumar2013decoupled}, but differs with the name VBS-QLRO$(\pi/2)$
            used by Sandvik. Further more, the phase boundary between the decoupled
            phase and the dimerized incommensurate phase is qualitatively different
            from both previous works.  We will prove this later. 
            For the time being,
            let us discuss the development of the dimerized and commensurate phase
            of the \jj model under the influence of LRI.  Sandvik shows by
            excitation state level crossing  that the dimerization critical point
            develops to (1.7, 0.41) as $\alpha$ reduces and then to a first order
            transition \cite{sandvik2010ground}. Below, we determine the development
            of the \MG point as well, but from an entanglement perspective. 
            }
        In the center of the phase diagram is a $\lambda$-shaped phase
        boundaries, with  the dimerized and commensurate phase beneath them. 
        Sandvik\cite{sandvik2010ground} has successfully determined the
        development of dimerization point of the \jj model
        to a multi-critical point at around (1.7, 0.41), and then to a first order
        phase transition, using level crossing. 
        Below, we study these phase boundaries again but from an entanglement
        perspective,  and we determine the development
        of the \MG point 
        \footnote { The starting point of this phase boundary 
        chose by Sandvik is $(\infty, 0.52)$, while we use $(\infty, 0.5)$. 
        This is a matter of angle viewing the commensurate/incommensurate transition
         either from momentum space or real space. This is discussed in detail
         in \onlinecite{bursill1995numerical}.  } as well.

        \begin{figure}   
            \centering
            \scalebox{0.55}[0.55]{\includegraphics{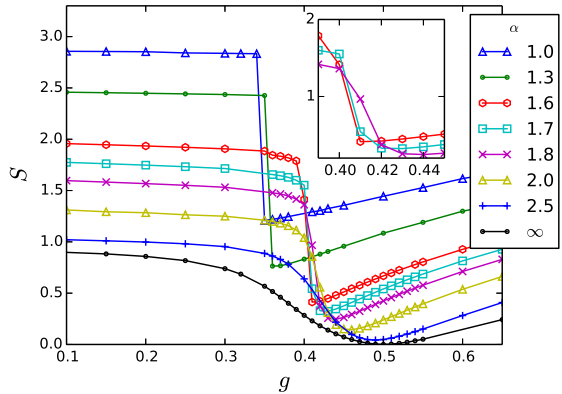}}
            \caption{\label{fig:lambad:boundary} Dependence of half chain
            entanglement $S_{N/2}$ on $g$ for several values of $\alpha$ with $N=100$.
            Inset displays in great detail of the lines
            around $g=0.4$  for $\alpha=1.6, 1.7$ and 1.8.}
            \end{figure}
        \comment{
            Instead, we first confirm one main result of \ref{sandvik2010ground},
            which also serve as a validation of the MPS approach. 
            In \ref{sandvik2010ground} one main result is, through the study of
            energy levels, the evolve of the $()$
            points and merge at a multiple critical point and then form a first
            order phase transition, altogether form a $\lambda$-shaped boundaries. We confirm
            this first result, but from the entanglement perspective, which also
            serve as a validation of the MPS approach. }
        Fig.\ref{fig:lambad:boundary} shows in particular dependence of
        entanglement on $g$ for several values of $\alpha$.  At $\alpha=\infty$,
        there are two turning points for the curve:  The first one is 
        related to the dimerization transition point at $g_c\simeq0.2411$ (where a
        gap opens and entanglement drops), but it is difficult to locate $g_c$ 
        accurately from entanglement; The other one is the minimum of
        entanglement just at the \MG  point $g_{MG}=0.5$. 
        The two points are smoothly connected and approaches as
        $\alpha$ reduces until the point $(1.7, 0.41)$
        (see inset of Fig.\ref{fig:lambad:boundary}), indicating shrinking of the
        \vbspi phase and finally terminating at that point. That multi-critical
        point is in agreement with the ED result obtained by extrapolation \cite{sandvik2010ground}.  
        \comment{
             With decreasing $\alpha$ these two quantum phase
            transition points approaches and merged into one at the point $(1.7,
            0.41)$ (see inset of Fig.\ref{fig:lambad:boundary}.  }
        The trajectory of $g_{MG}$ as $\alpha$ changes appears as a valley in
        the entanglement plane of 
        Fig.\ref{fig:EE}, where it should also have minimum correlation length,
        and thus  can be thought of as a disorder-line\cite{chitra1995density}, separating
        phases with commensurate and incommensurate correlation on either side. After the
        multi-critical point, entanglement becomes discontinuous displaying a
        sudden deep drop, which clearly marks the first order phase transition from the 
        AFM phase to the \qlropihalf phase \cite{sandvik2010ground}.

        \begin{figure}   
            \centering
            \scalebox{0.55}[0.55]{\includegraphics{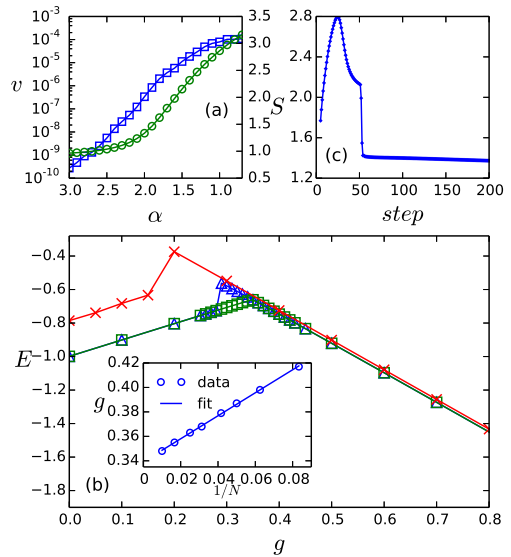}}
            \caption{\label{fig:energy} 
            (a) Dependences of variance (open square)
            and middle chain entanglement (open circle) on $\alpha$ shown
            together for comparison for fixed $g=0.0$
            and $N=100$, $D=520$. 
            (b) Ground state energy as a function of $g$ at $\alpha=1.0$
            obtained from different algorithms. Open
            triangle: VMPS initiated from iDMRG or random state; open square: 
            VMPS with proper initial state; cross: iDMRG. 
            Inset shows extrapolation of position of peak of energy using $N$ ranging from 16
            to 100. The fitting line is $g=1.0/N+0.339$.  
            (c) Middle chain entanglement as a function of iDMRG steps at
            $\alpha=1.0, g=0.0$ with $D=1000$.  }
            \end{figure}
        
        The rather high amount of entanglement in the AFM phase poses
        considerable difficulties to the MPS method, since the computational
        effort for it scales exponentially with entanglement
        \cite{vidal2003entanglement}. Here we would like to elaborate on these
        difficulties. Fig.\ref{fig:energy}(a) shows dependence of both average
        variance and entanglement on $\alpha$ at $g=0.0$.  This serves as a
        benchmark of the accuracy viable. One can see that variance increases
        radically with decreasing $\alpha$.  
        These restrict us to $\alpha\ge0.7$ for the variance smaller
        than 1e-4  for the maximal length $N=100$ and largest $D=520$ used.
        And we find much more sweeps (around 10 times) needed for convergence for small
        $\alpha$. 
        Further more, there is metastable state issue to the left of the
        first order transition point. As shown in
        Fig.\ref{fig:energy}(b), it is
        prone to get stuck on an excitation level which should have less entanglement than
        the ground state, if one uses random state or iDMRG for an initial
        state.  This leads to a wrong position of the peak of energy
        (the transition point) compared with ED\cite{sandvik2010ground}. For a given $N$, larger $D$
        can shrink the region being stuck, but soon become unpractical.  A
        two-site algorithm with density matrix correction won't solve it either.
        It turns out a nice solution is to provide a better initial state,
         e.g., use the state of a smaller $g$ as
        the input of larger $g$ close to the right of the boundary (see
        [\onlinecite{stoudenmire2012studying}] for alternative ways such as adding
        a pinning term for fixing this). In this way, the peaks for each lengths
        are unambiguously determined and the extrapolated value of the
        transition point 
        is $g_c=0.339$ (see inset of Fig.\ref{fig:energy}(b)). As for the infinite
        algorithm, the metastable issue is more severe. It is stuck in a wider
        range, which, we however haven't found a way to avoid.  
        For $g$ close to 0 the energy still deviates with VMPS. 
        This is not because of getting stuck but is a convergence problem due to too
        fast growing of entanglement and at the same time relatively slow convergence of
        energy. As shown in Fig.\ref{fig:energy}(c), entanglement
        suddenly drops after around 30 iteration steps (120 chain length) if
        $D=1000$ is kept not increased  and
        eventually converged to a wrong fixed point. One could stop
        iteration before the drop (this is where the data points of energy 
        we adopted), but energy and other quantity are far from
        convergence. In all we find good convergence of VMPS for the parameter
        range studied, while iDMRG has either metastable state or convergence problems
        in the AFM phase. 

\subsection{scaling of entanglement under LRI}  \label{sec:results:EE_scaling}
    \begin{figure}   
          \centering
          \scalebox{0.55}[0.55]{\includegraphics{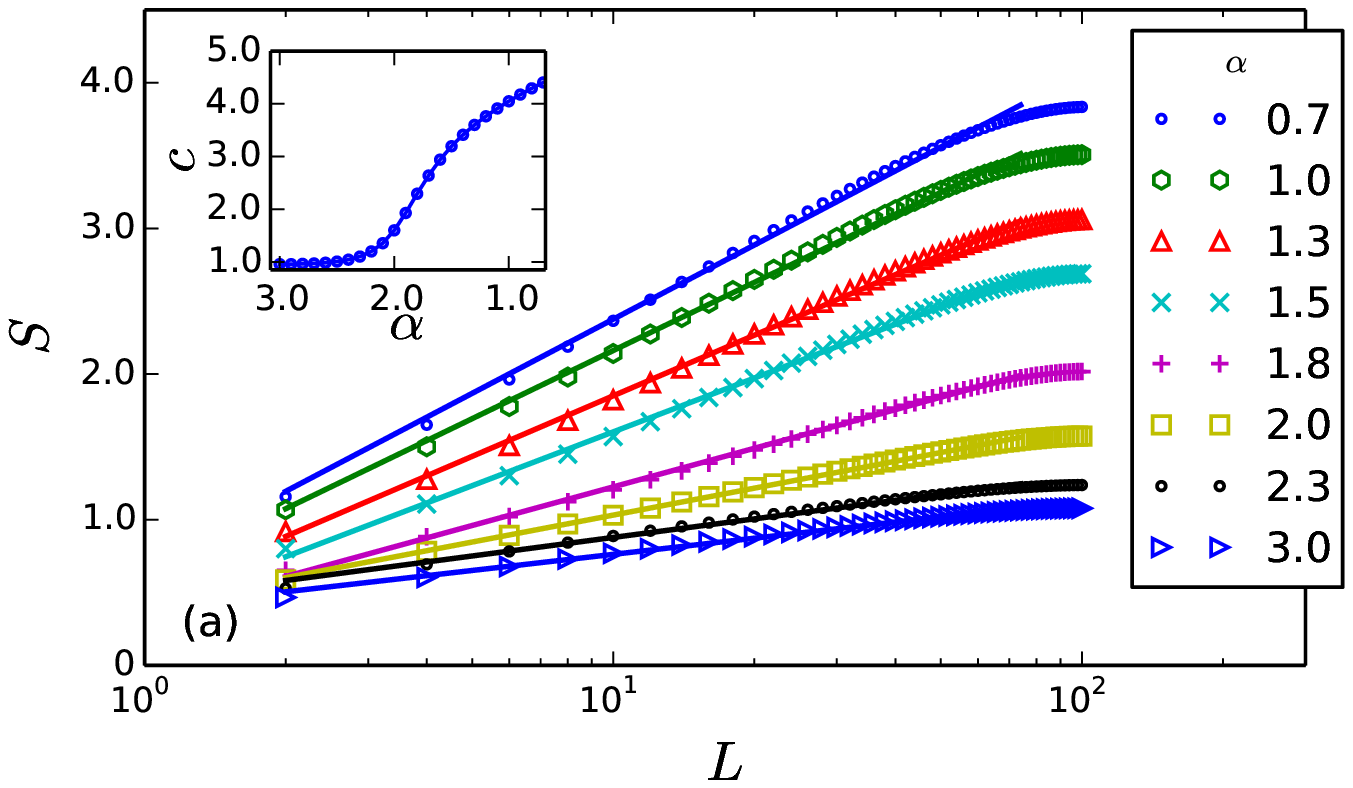}}
          \scalebox{0.55}[0.55]{\includegraphics{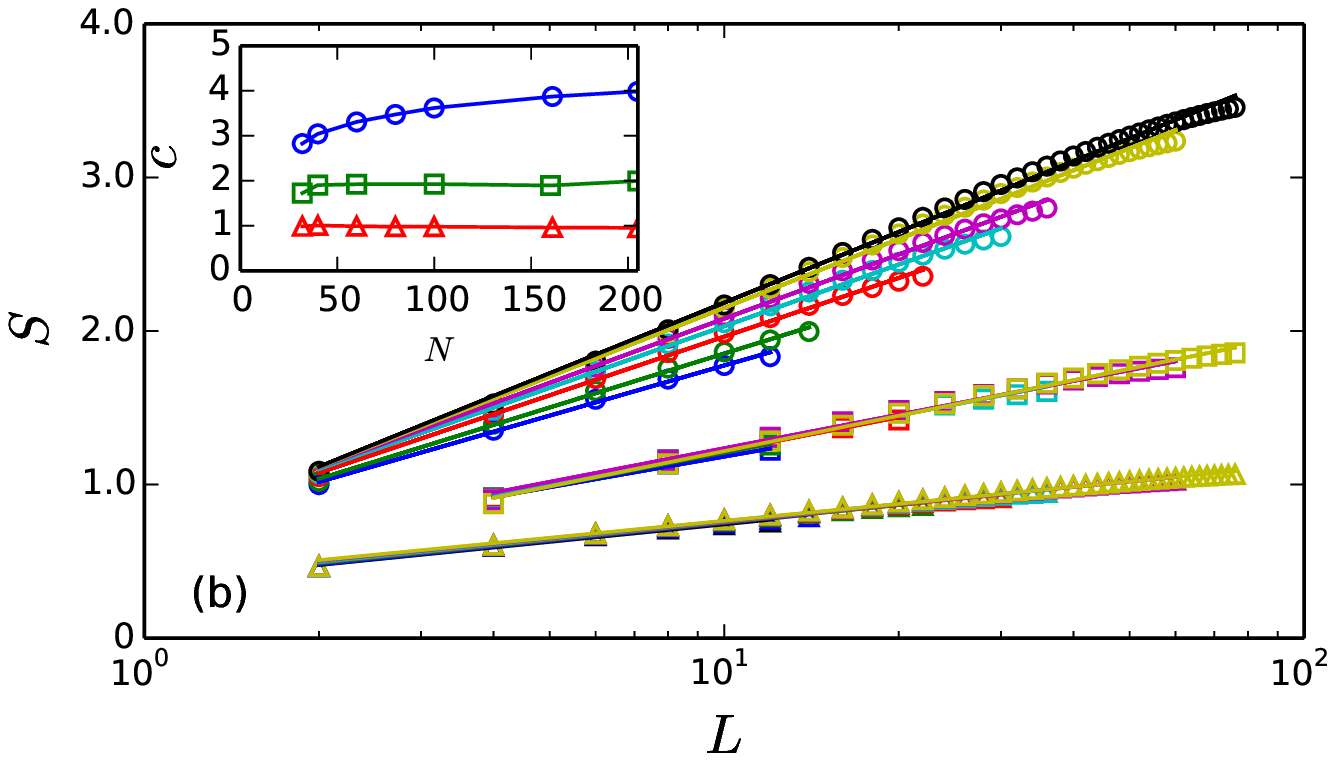}}
          \caption{\label{fig:EE:vs:L:N200} (a) Bipartite entanglement entropy
          $S$ as a function of subsystem length $L$ for each 
          $\alpha$ with $g=0.0$. Only bipartition on even bonds (even $L$) are
          drawn. Solid lines are fitting to $S=\frac{c}{6}\ln(L)+\text{const}$ for 
          $1\leq L\leq75$. Inset shows the extracted $c$ vs. $\alpha$. 
          (b) Three groups of curves  at $(1.0, 0.0)$ (open circle), (1.0,
          0.7) (open square) and (3.0, 0.0) (open triangle) each shows dependence of $S$
          on $L$ for different system sizes $N$ ranging from 32 to 200. 
          $L$ is restricted to no larger than $\frac{3}{8}N$, i.e. not close the
          chain center. Solid 
          lines are fitting to the function as in subgraph (a). Inset shows $c$ extracted
          from fitting of each group of curves  as a function of $N$.  } 
      \end{figure}

    As shown above, the highest amount of entanglement is
    found when frustration is zero and $\alpha$ is small.
    \comment{
        focusing on the line of $g=0.0$. 
        We study for each $\alpha$ the dependence of entanglement on subsystem lengths.
    }
    At $g=0.0$, according to previous ED \cite{sandvik2010ground}
    and quantum Monte Carlo \cite{laflorencie2005critical} studies, the system
    undergoes a continuous phase transition from the \qlropi to AFM phase at
    $\alpha_c\simeq2.22$.  
    Recall that for a conformal invariant system with open boundary condition, the entanglement scales
    logarithmically with subsystem length \cite{vidal2003entanglement, * calabrese2004entanglement, * holzhey1994geometric} 
    \begin{equation} \label{eq:ee:vs:l} S
        \sim \frac{c}{6}\ln(L), \end{equation} 
    where $c$ in the prefactor is
    identified with the central charge of the relevant field theory.  The QLRO
    phase belongs to this category and has $c=1$. 
    We expect changing of behaviour of $S(L)$ around
    $\alpha_c$, and wonder to what extent the entanglement in the AFM phase with strong LRI 
    violates the area law.    
    \comment{
        Beside, as computational cost is rather hight in the AFM phase, VMPS is
        constrained to $N\leq100$. While use iDMRG to generate finite open chains
        for $100<N\leq200$.}
    
    To this end, we focus primarily on the line of $g=0.0$ and use both VMPS and
    iDMRG to simulate various sized systems. The former, being free of environment 
    error \cite{schollwock2005density} and better controlled, is used for
    $N\leq100$, while the later having reduced computation cost, is used for
    $100<N\leq200$ and validated by the consistency with small
    chains. We first present $S(L)$ for the largest size $N=200$ for each
    $\alpha$ in Fig.\ref{fig:EE:vs:L:N200}(a).  \comment{
        and reveal the salient finite size effect on $S(L)$ for various size $N$
        in Fig.???. }
    One can see remarkably that, for all $\alpha$ values $S$
    scales approximately logarithmically with $L$ (for $L$ not
    too close to the chain center), and that $S(L)$ increases clearly faster
    for smaller $\alpha$. To quantify and compare them,  it is tempting to use
    Eq.\ref{eq:ee:vs:l} to fit each curve and extract a value of $c$. We may call $c$
    generally an ``effective central charge'', as the AFM phase is not conformal
    invariant \cite{laflorencie2005critical}.   
    As shown in the inset of Fig.\ref{fig:EE:vs:L:N200}(a), $c$ is near to 1.0
    for $\alpha\leq2.5$, after that, it increases quickly with $\alpha$ and
    reaches a large value.  This behaviour is in overall agreement with the
    transition point $\alpha_c$. 

    The above extracted values of $c$ in the AFM phase are, however, not quite validated.
    For one thing, closely examining the fitting, one finds small deviations
    from perfect logarithm: $S(L)$ seemingly increases slightly faster for
    larger $L$. For another, and more significantly, 
    $S(L)$ actually has a salient dependence on the total system size, as a
    result the $c$ values are only specific to $N=200$. 
    To manifest this finite size effect on $S(L)$, we compare three points (1.0,
    0.0), (1.0, 0.7) and (3.0, 0.0) which are respectively representatives of
    the AFM, decoupled, and QLRO phases. For each point we plot $S(L)$ for various system sizes,
    as shown in Fig\ref{fig:EE:vs:L:N200}(b). Since they all have
    (approximately) logarithm divergence, the values of $c$ are extracted for
    each $N$, and shown in the inset of the graph. One can find that 
    a clear dependence of $S(L)$ and $c$ on $N$ is unique to the AFM phase. (In
    the decoupled phase, $c$ are close to 2.0, which is expected. Because, as will be
    shown later, it is a system of two approximate Heisenberg chain, whose
    central charge is just the sum of each one's.) 
    The dependence of $S(L)$ on $N$ implies that the maximal size $N=200$ should
    be not enough. We will try to give an extrapolated result for it later, but
    below we first try to interpret the results obtained.  
    
    So far in the above we observed that for all $\alpha$ values, $S(L)$ have
    approximately logarithm dependence, while the slope $dS/d(ln(L))$ i)
    apparently increases for smaller $\alpha$, ii) slightly increases for larger
    $L$ and iii) increases for larger $N$. We give a naive explanation for these 
    from a valence-bond description of the singlet ground state of
    antiferromagnet. 
    A valence-bond is a singlet pair $\frac{1}{\sqrt{2}}|\uparrow_i \downarrow_j -\downarrow_i \uparrow_j
    \rangle$, where $i$ and $j$ are on opposite sublattices of of a bipartite
    lattice.     
    It is known that, a SU(2) singlet ground state can always be
    represented in a valence-bond basis which is all possible covering of
    singlet pairs on the chain $[1,N]$ for even $N$.  
    Each singlet pair is maximally entangled, and the value is $\ln(2)$. Thus
    bipartite entanglement can be measured as the number of bonds cut by the
    bipartition  times $\ln(2)$ \cite{alet2010valence,*
    chhajlany2007topological}.  This give a appealing geometrical interpretation of the ground
    state entanglement. For the
    Heisenberg model, $\vec S_i\cdot\vec S_{i+1}$ favors forming of singlet between
    spin $i$ and $i+1$, but many body effects eventually leads to forming of
    complex distribution of configurations of valence-bond
    \cite{chhajlany2007topological} --- including 
    pairs apart in arbitrarily long distances. Note it is the long distance
    entangled pairs that lead to divergence of entanglement with increasing $L$. 
    Notably, for the unfrustrated regime of the Hamiltonian \ref{eq:ham},  
    long-range terms $J_{j-i}\vec S_i\cdot\vec S_j$ (we may call them bonds of the
    Hamiltonian) favor directly long distance singlet pairs. 
    It is then reasonable to assume that the number of valence-bond in the
    ground state across $[1,L]$ and $[L+1,N]$ is in a way positively correlated with the
    sum of the strength of all bonds of the Hamiltonian across them, i.e. with
    the quantity
    \begin{equation}
        \label{eq:JL}
        {\mathcal{J}} \equiv \sum\nolimits_{1 \le i \le L < j \le N} {{{(j -
        i)}^{ - \alpha }}}. 
    \end{equation}
    (Note that we can safely ignore the normalization in the
    Hamiltonian when $g=0$.)
    Based on this, we may understand the properties just stated. For i), \JL
    increases with smaller $\alpha$, so entanglement increases faster as well.
    (But this does not explain why they are close to logarithm for all the
    $\alpha$); 
    For ii) and iii), \JL sums over $L\times(N-L)$ (approximately $L\times N$ for large $N$
    and small $L$) bonds, while in contrast, for the QLRO phase which is
    essentially short-ranged, there is
    always only one bond connects the two regions irrespective of the position of
    the cut or the system size. This helps to understand why in the AFM phase
    the approximate logarithm function $S(L)$ increases even faster when $L$ or
    $N$ expands, while it does not in the QLRO phase.

     \begin{figure}   
          \centering
          \scalebox{0.55}[0.55]{\includegraphics{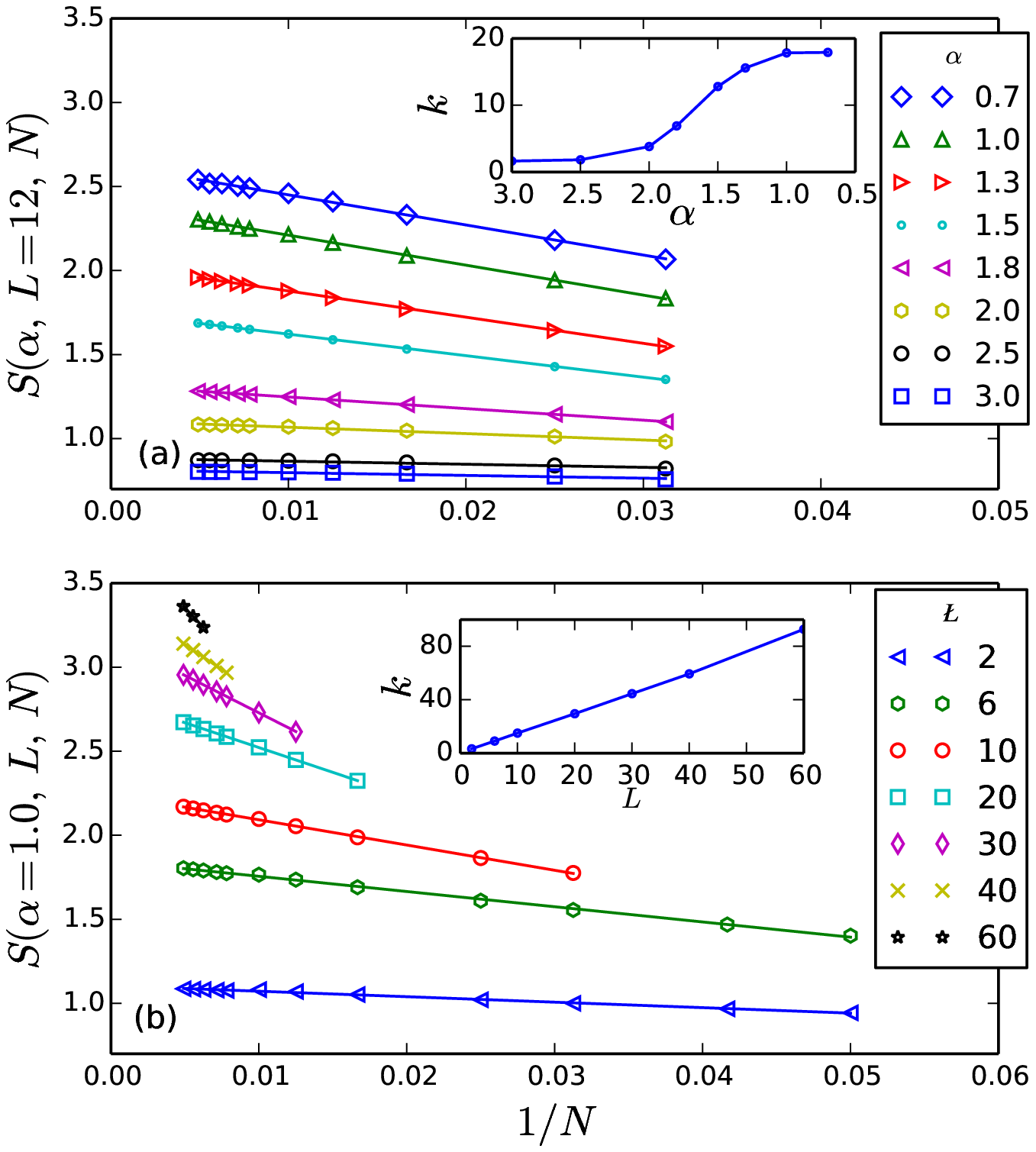}}
          \scalebox{0.55}[0.55]{\includegraphics{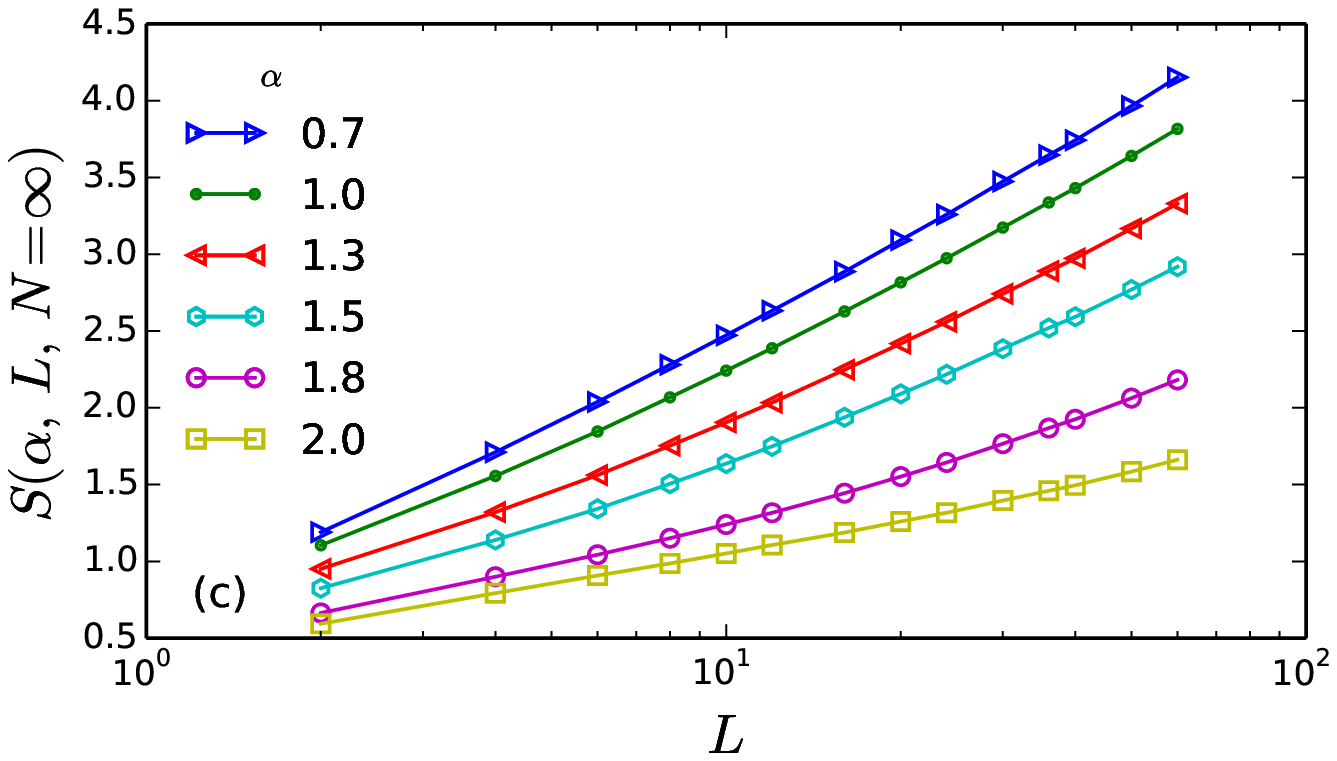}}
          \caption{\label{fig:EE:vs:L:Ninf} (a) $S(\alpha,L,N)$ versus inverse
          chain length for different $\alpha$ and fixed $L=12$. Solid
          lines are fitting to $S(\alpha,L,N)=-k/N + S(\alpha,L,\infty)$ which
          is a linear function of $1/N$. (b)
          $S(\alpha,L,N)$ versus inverse
          chain length for different $L$ and $\alpha=1.0$. 
          Solid lines are fitting to the same function as in subgraph (a).          
          The system sizes $N$ used are restricted to $[N_{min},200]$, with 
          $N_{min}$ depending on $L$, such that $\frac{3}{8}N_{min}\geq L$, namely,
          $L$ should not close to the chain center. 
          (c) Extrapolated $S(\alpha, L, N)$ value to $N=\infty$ as a function
          of $L$ ($L\leq 64$) for several $\alpha$,  obtained by curve fitting as above. The
          lines are guide for eyes. 
          }
          \end{figure}
    
    \comment{
        The dependence of $S(L)$ on $N$ implies that the maximal size $N=200$ should
        be not enough. }
    To get a glance of $S(L)$ in he thermodynamical
    limit, we try to extrapolate the values $S(\alpha,L, N)$ in $N$ for
    each fixed $\alpha$ and $L$. Hereafter we make the dependence of $S$ on
    $\alpha$, $L$ and $N$ explicit, leaving fixed $g=0.0$ implicit. 
    In Fig.\ref{fig:EE:vs:L:Ninf}(a) we fix $L=12$ and show for different $\alpha$ the dependence
    of $S(\alpha,L,N$) on $1/N$. In Fig.\ref{fig:EE:vs:L:Ninf}(b), we fix
    $\alpha=1.0$ and show for different
    $L$ the dependence of $S(\alpha, L,N)$ on $1/N$. They  
    show that the extrapolation is plausible. For different $L$ and
    $\alpha$ there is always a $1/N$ relation 
    \begin{equation}
        \label{eq:sl:extrapolate}
        S(\alpha,L,N) = -k\frac{1}{N} + S(\alpha,L,\infty), 
    \end{equation}
    but with a non-travail coefficient $k$ depending on $\alpha$ and $L$.  Two 
    profiles of $k(\alpha,L)$ for fixed $L=12$ or $\alpha=1.0$ are shown in the
    insets of Fig.\ref{fig:EE:vs:L:Ninf}(a) and (b).   
    The finite size effect is more salient for smaller $\alpha$ or larger $L$. This is
    in agreement with the above argument from the valence-bond description. 
    
    Then extrapolated $S(\alpha, L, \infty)$ for more $\alpha$ and $L$
    combinations are shown in Fig.\ref{fig:EE:vs:L:Ninf}(c). Note that it should
    be understood as a result for semi-infinite chain $[1, \infty)$. 
    One can see that each curve is still close to a logarithm
    function, with the slop $dS/d(ln(L))$ 
    slightly increases with larger $L$.  The maximal $L$ evaluated is rather
    restricted, it is not very clear whether $dS/d(ln(L))$ keeps increasing
    slowly or approaches a constant for even larger $L$. For the former, the function
    form can be e.g. $S=a_1\ln(L)^2+a_2\ln(L)+a_3$, while for the later, it can be
    e.g. $S=a_1ln(L)+a_2/L+a_3$. We have verified that both can give a good fit for the
    curves (not shown), other than a simple logarithm function. But in view of
    the $1/N$ correction in Eq.\ref{eq:sl:extrapolate}, and that in
    Fig.\ref{fig:EE:vs:L:N200}(b) curves for different $N$ are all close to 
    logarithm functions, we are prone to the later.

    

\subsection{phase boundary at moderate frustration} \label{sec:resuts:moderate_g}
        \begin{figure}   
              \centering
              \scalebox{0.55}[0.55]{\includegraphics{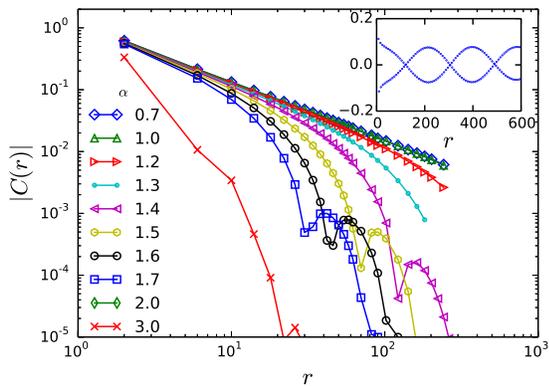}}
              \caption{\label{fig:corr} Absolute value of the intra sublattice correlation $C(r)$
              for each $\alpha$ at $g=0.7$. 
              $C(r)$ are obtained by evaluating Eq.\ref{eq:corr} upon IMPS from 
              iDMRG simulations with $D$ up to 960. Inset shows 
              $C(r)$ at $\alpha=1.4$ and $g=0.7$ multiplied by
              $\sqrt{r}e^{r/\xi}$ with $\xi=44.7$ using iDMRG and $D=640$.             
              }
              \end{figure}

        We next turn to the right part of the phase diagram.  
        Focusing on one line $g=0.7$, we measure the spin correlation 
        \begin{equation}
            \label{eq:corr}
            C(r) = \langle S_i^zS_{i + r}^z\rangle  
            \end{equation}       
        and see how it changes as $\alpha$ reduces. 
        Here the spin chain is considered to
        be divided into odd and even sublattices, as is usually did for the
        \jj model. Fig.\ref{fig:corr} shows correlation between spins within a 
        same sublattice (even $r$).   
        One can see that for large $\alpha$ values $C(r)$ decreases exponentially with jumps in
        the curve. 
        The jumps signify the incommensurate behaviour: Following the
        treatment of White and Affleck \cite{white1996dimerization} of the \jj
        model, we multiplying $C(r)$ e.g. at $\alpha=1.4$ and
        $g=0.7$ by $\sqrt{r}e^{r/\xi}$, then the
        sinusoidal modulation is clearly seen in the inset of the graph, where  
        the correlation length $\xi=44.7$ is
        chosen such that the beats of the amplitude are as flat as possible. 
        By evaluating $C(r)$ for $r$ up to 1000, we only find jumps for
        $\alpha>1.2$. 
        For $\alpha\leq1.2$,
        $C(r)$ displays algebraical decay (for reference $C(r)\sim 1/r^\gamma$ with 
        $\gamma=1.18$ at $\alpha=1.0$). This indicates a possible 
        critical value $\alpha_c$ of a continuous phase transition at roughly 1.2. 
        
        \begin{figure}   
              \centering
              \scalebox{0.55}[0.55]{\includegraphics{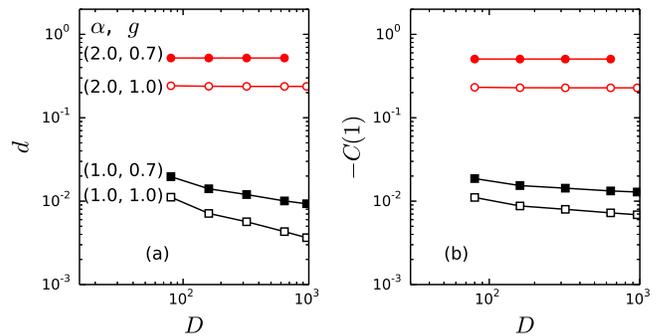}}
              \caption{\label{fig:dimer} 
              Left and right shows respectively  
               dimerization $d$ and minus of inter sublattice correlation for
              $r=1$, measured upon IMPS, as a function of iDMRG truncation
              dimension $D$ for several $(\alpha,g)$ tuples.}
              \end{figure}
        
        \comment{
            In previous  studies with short chains, Sandvik
            \cite{sandvik2010ground}  estimates that the transition should be from
            the VBS phase to a coexisting VBS+QLRO($\pi/2$) phase, while Kumar and
            Soos \cite{kumar2013decoupled} predict it VBS phase to a decoupled
            phase.  
            We will show below that for $\alpha<\alpha_c$ there should be no 
            VBS order, thus we agree with Kumar and Soos that the system is
            essentially decoupled into two critical spin chains. 
            But there is still a crucial difference from Kumar and Soos. The phase
            boundary estimated by them is starting from a unique multi-critical at
            around (1.7, 0.41) and approaches $\alpha=\infty$ as $g$ increases. The
            starting point is incorrect.  Because as shown above, the critical point
            (1.2, 0.7) is already above that point, so the VBS(incomm.)to decoupled
            boundary should intersect the first order transition at a second
            multi-critical point above the first one (see
            Fig.\ref{fig:phase:diagram}).  The significance of this is that there
            indeed can be direct transition from AFM long-range order to VBS order
            \cite{sandvik2010ground}. 
            }

        In previous studies with short chains, Sandvik
        \cite{sandvik2010ground}  estimates that the transition should be from
        the VBS phase to a coexisting VBS+QLRO($\pi/2$) phase, while Kumar and
        Soos \cite{kumar2013decoupled} predict that the transition is from VBS phase
        to a decoupled phase.  We agree with Kumar and Soos that for small
        $\alpha$ there should be no VBS order and is essentially decoupled, but
        from an independent and more direct way: In fact the correlation shown above already
        implies no VBS order for $\alpha<\alpha_c$, because the
        presence of incommensurate behaviour is related to a finite dimerization
        \cite{white1996dimerization}. We also show directly the dimerization
        order parameter \cite{bursill1995numerical,white1996dimerization} $d =
        \langle S_i^zS_{i + 1}^z\rangle  - \langle S_i^zS_{i - 1}^z\rangle,$ for
        several $(\alpha,g)$ tuples in
        Fig.\ref{fig:dimer}(a).  One can see that, for $\alpha=2.0>\alpha_c$, $d$ are 
        clearly nonzero; While for $\alpha=1.0<\alpha_c$,  $d$ are very small and
        appear to vanish in the limit $D\to\infty$. In addition, the inter-chain
        correlation is very small for $\alpha<\alpha_c$, as shown in
        Fig.\ref{fig:dimer}(b).
        (Here it suffices to consider $C(r)$ for $r=1$, since $C(r)$ further
        decays for larger odd $r$.) 

        The transition points for other $g$ can be determined  likewise  and we
        find that $\alpha_c$ increases with $g$, and it should be that
        $\alpha_c\to\infty$ as $g\to\infty$. This gives a approximate phase
        boundary between VBS(incomm.) and the decoupled phase in
        Fig.\ref{fig:phase:diagram}. Note crucially that the starting point of the phase
        boundary is distinct from both Kumar and Soos's or Sandvik's result, in
        which, it starts at a unique multi-critical at around (1.7, 0.41). The
        significance of this is that there indeed can be direct transition from
        AFM long-range order to VBS order \cite{sandvik2010ground}.
        
        The phase transition can be understood as follows: At moderate or large
        $g$,  as $\alpha$ reduces, the couplings $J_{r}$ for $r\neq2$ (including
        $J_1$) all become very small due to a large normalization factor, and
        eventually the next-nearest neighbour term $g\vec
        S_{i} \cdot \vec S_{i+2}$ dominates.  The later induce a background of
        antiferromagnetic order on either sublattices, this amounting to a period 4
        structure in the entire lattice.  It is not difficult to see that,
        the rest enormous but small couplings $J_{r\neq2}$ now 
        have conflicting signs against this background period, and their
        effects should be largely smeared out.  So for mall enough $\alpha$ it
        is decoupled into nearest-neighboured
        Heisenberg model with coupling $g$ and with long-range terms as
        perturbations.  Note that for the \jj model ($\alpha=\infty$), White and Affleck
        \cite{white1996dimerization} used
        field theory to predict that there is exponential small gap and
        dimerization for arbitrary large $g$, except for $g=\infty$ where it
        decouples exactly into two Heisenberg chain,  and supported it by DMRG.
        While our arguments above  essentially states that, at small $\alpha$,
        the spin chain can be decoupled for modest $g$.  Nevertheless, the
        evaluation of $C(r)$ is still not very long and  the decay of  $d$ with
        $D$ is somewhat slow, so we are still not completely sure whether there
        are incommensurate modulation with very long period or dimerization
        (and also spin gap) should be exponentially small but nonzero even as
        $\alpha$ approaches 0, which is very difficult to confirm numerically.
        A field theory study for small $\alpha$ may be desirable as well as that
        for the \jj model at large $g$.

\section{conclusion}  \label{sec:conclusion}
    
    In summary, we studied the frustrated spin chain with long-range
    interactions using the matrix product state approaches. We found that it is
    most entangled in the rotation invariant AFM long-range ordered phase, where
    the entanglement scales approximately in logarithmic form.  But we miss a
    complete understanding for the logarithmic scaling. The maximal systems size
    studied is not large and $\alpha$ is still not close to 0. It is worthwhile
    to check the persistence of the logarithmic form in the asymptotic scaling
    of entanglement for large subsystem size and smaller $\alpha$ in future
    works.  We also studied correlation and dimerization for moderate
    frustration and determined an approximate boundary for the transition from
    the dimerized and incommensurate phase of the \jj model to a decoupled
    phase.  

    Our work implies that in an antiferromagnetic spin chain with LRI,
    frustration terms usually hamper long-range entanglement. Consider a series
    of power law decayed models $H(\{s\})=\sum\nolimits_{i,r} {s(r)r{^{-\alpha}
    }{{\vec S}_i}\cdot{{\vec S}_{i+r}}}$, with the signs $s(1)=+1$  while
    $s(r>1)\in \{+1, -1\}$ indeterminate.  At small $\alpha$, their entanglement
    should be bounded by the unique nonfrustrated one i.e. $s(r)=(-1)^{r-1}$,
    where perfect AFM long-range order is formed. 
    
    The scaling of entanglement shown in this work and previous works
    \cite{latorre2005entanglement,koffel2012entanglement} indicates that,
    although LRI can usually increase entanglement, but it not necessarily leads
    to large entanglement and severe violation of area law. 
    The preservation or (severe) violation of area law may be a joint
    effects of interaction range, symmetry constrains (e.g.  translation
    invariance), ground state degeneracy, and etc, which remains to be
    clarified. 
    
    \comment{
        By large entanglement, one may refer to, say in 1D e.g. a power scaling
        $S\sim L^x$ with $0<x\leq1$.  In contrast, there are examples of
        nearest-neighboured models that can have $\sim L^{\frac{1}{2}}$
        \cite{movassagh2014power} or even linear \cite{vitagliano2010volume} scaling
        of entanglement. But is also examples that 
        The preservation or (severe) violation of area law may be a joint
        effects of interaction range, symmetry constrains (e.g.  translation
        invariance), ground state degeneracy, and etc, which remains to be
        clarified. 
        }

    \comment{
        The entanglement in the AFM phase should be a upper bound for
        all 1D spin chains models with arbitrary power law decay long range and two
        body interaction. Because, first, models with rotation invariance is generally
        more entangled than polarized ones; Second, the rotation invariant model
        \ref{eq:ham} with a frustration term is illustrative enough to show that any
        change in the Hamiltonian that breaks the perfect long range order should
        only reduce entanglement. So it is reasonable to conjecture that all such
        type of systems should have at most logarithmic divergence of entanglement,
        and is in this sense not severely breaking the area law. 
        }
   
\section*{acknowledgement}    
     ZHL thanks A.W. Sandvik and M.Q. Weng for valuable discussions.  
     We acknowledge an anonymous referee whose comment induced us to do a
     systematic analysis on the finite size effect on entanglement scaling. 
     This work is supported by National Natural Science Foundation of China under Grant 
     No. 11375168.  The simulation is mainly conducted on the Supercomputing
     system in the Supercomputing Center of USTC.  
     
    

    \section*{References}
    \bibliography{mera-long-range}

 

\end{document}